\documentclass[12pt]{iopart}
\pdfoutput=1
\usepackage{graphicx}
\usepackage{fancyhdr}
\usepackage{amssymb,amsfonts,color}

\begin{document}

\title[]{Towards the map of quantum gravity}

\author{Jakub Mielczarek$^{*}$ and Tomasz Trze\'{s}niewski$^{*,\dagger}$ }

\address{${}^*$Institute of Physics, Jagiellonian University, {\L}ojasiewicza 11, 30-348 Krak\'{o}w, Poland \\
${}^{\dagger}$Institute for Theoretical Physics, University of Wroc{\l}aw, Pl.\ Borna 9, 50-204 Wroc\l{}aw, Poland}

\begin{abstract}
In this paper we point out some possible links between different approaches to quantum gravity and theories of 
the Planck scale physics. In particular, connections between Loop Quantum Gravity, Causal Dynamical 
Triangulations, Ho\v{r}ava-Lifshitz gravity, Asymptotic Safety scenario, Quantum Graphity, deformations of 
relativistic symmetries and nonlinear phase space models are discussed. The main focus is on quantum 
deformations of the Hypersurface Deformations Algebra and Poincar\'{e} algebra, nonlinear structure of phase 
space, the running dimension of spacetime and nontrivial phase diagram of quantum gravity. We present an 
attempt to arrange the observed relations in the form of a graph, highlighting different aspects of quantum 
gravity. The analysis is performed in the spirit of a mind map, which represents the architectural approach 
to the studied theory, being a natural way to describe the properties of a complex system. We hope that the 
constructed graphs (maps) will turn out to be helpful in uncovering the global picture of quantum gravity as 
a particular complex system and serve as a useful guide for the researchers.
\end{abstract}

\maketitle

\section{Introduction}

\begin{figure}
\begin{center}
\includegraphics[width=16cm]{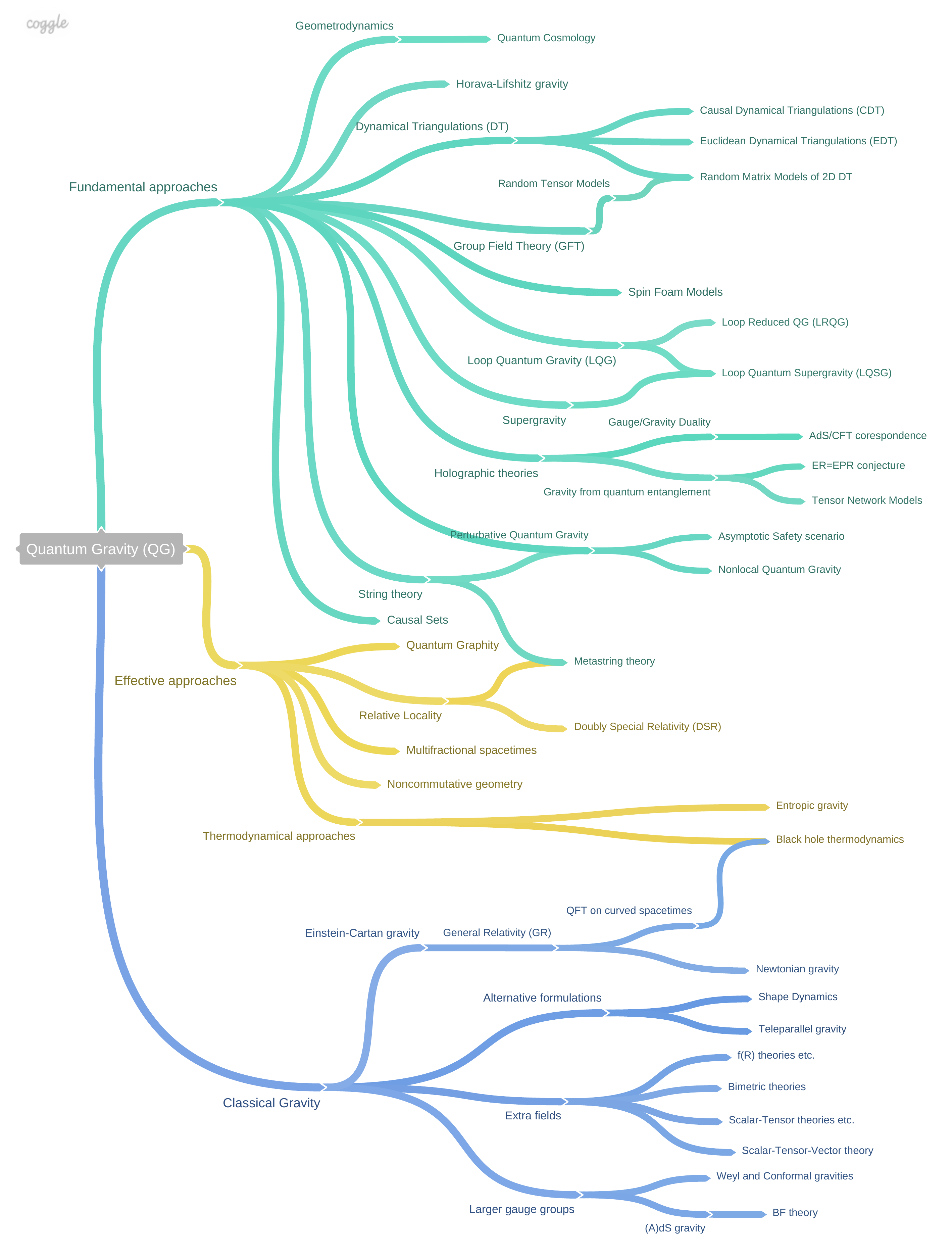}
\end{center}
\caption{A simplified taxonomy of approaches to quantum gravity.}
\label{Fig1}
\end{figure}

For around 100 years\footnote{The importance of quantum effects in gravity was noted by Einstein already 
in 1916 \cite{Rovelli:2000aw}.} of constantly branching research, the landscape of quantum gravity 
investigations has become very broad and diverse. Nowadays it contains many scattered building blocks -- 
conceptual dilemmas and conjectures, novel mathematical tools, models at the nascent stage of development 
and theories that are already quite extended. The usual perspective is that we have distinct, competing 
approaches to the construction of quantum gravity as well as the modification of classical general 
relativity (assuming that the latter is a useful intermediate step), which arose partly due to the 
scarcity of experimental guidelines. They can be more or less sorted in several ways, e.g. some formulations 
of the quantum theory are classified as canonical and some as covariant, depending on the applied 
procedure of quantization. One possible realization of such a taxonomy of (modified) classical and 
quantum gravity has been depicted in Fig.~\ref{Fig1} but it should be understood as an outline rather than a precise diagram. In the proposed scheme we roughly divide 
quantum gravity approaches and the more general theoretical frameworks according to whether they are built around the assumed fundamental 
degrees of freedom or represent some type of the effective description (at least at their current 
stage of development). There is also included a spectrum of potential classical approximations to the theory, consisting of general relativity and its various modifications, 
which are extensively reviewed in e.g. \cite{Blagojevic:2012gn,Clifton:2012my}. Certain frameworks can be seen as reductions or developments of others, sometimes due to the synergy of different branches of research, but for simplicity we show only a few links (and we do not include different versions of quantum cosmology). In general, we have restricted here to the approaches 
that are either most popular or especially interesting in the context of issues that will be considered 
in this paper. However, simple taxonomy can not help to address the problem of consistency or contradiction 
between various existing models. 

On the other hand, as we will argue in the next sections, some sort of convergence in physical predictions 
of various frameworks has recently started to become apparent. If a lucky coincidence is excluded, this may 
indicate significant underlying connections. Moreover, there is an interesting idea to consider 
(see especially Sec.~\ref{sec:phases} and references therein) that (quantum) gravity is actually, or at least can be treated as, a complex 
system, which exhibits emergent phenomena. Its fundamental degrees of freedom might behave in a 
very different way than the gravitational field at some effective level. Then individual models could be 
potentially applied to different layers of the theory and in this sense become unified. In order to explore 
all of these possibilities we should employ a more architectural way of thinking, by which we mean trying 
to uncover the hypothetical structure of the still incomplete theory and to understand its internal functions. 
This contrasts with what may be called the engineering approach, in which independent groups 
of researchers keep working on their rival theories and which has dominated quantum gravity over the years. 
In our opinion, only an appropriate combination of both architecture and engineering is able to bring us 
to the destination point. 

The architecture considered in this paper is not a very sophisticated one but every story has its beginning. 
In order to represent the structure and phenomena of the discussed theory we will apply the idea of a mind 
map. The nodes of such a graph (playing the role of bricks) will denote different models of quantum gravity 
as well as their features and known results. The links (which are like mortar) will tell us which of these 
elements are directly interrelated. The arrow of a link will not mean an implication but show a direction 
in which the graph should be read. 

\begin{figure}
\begin{center}
\includegraphics[width=16cm]{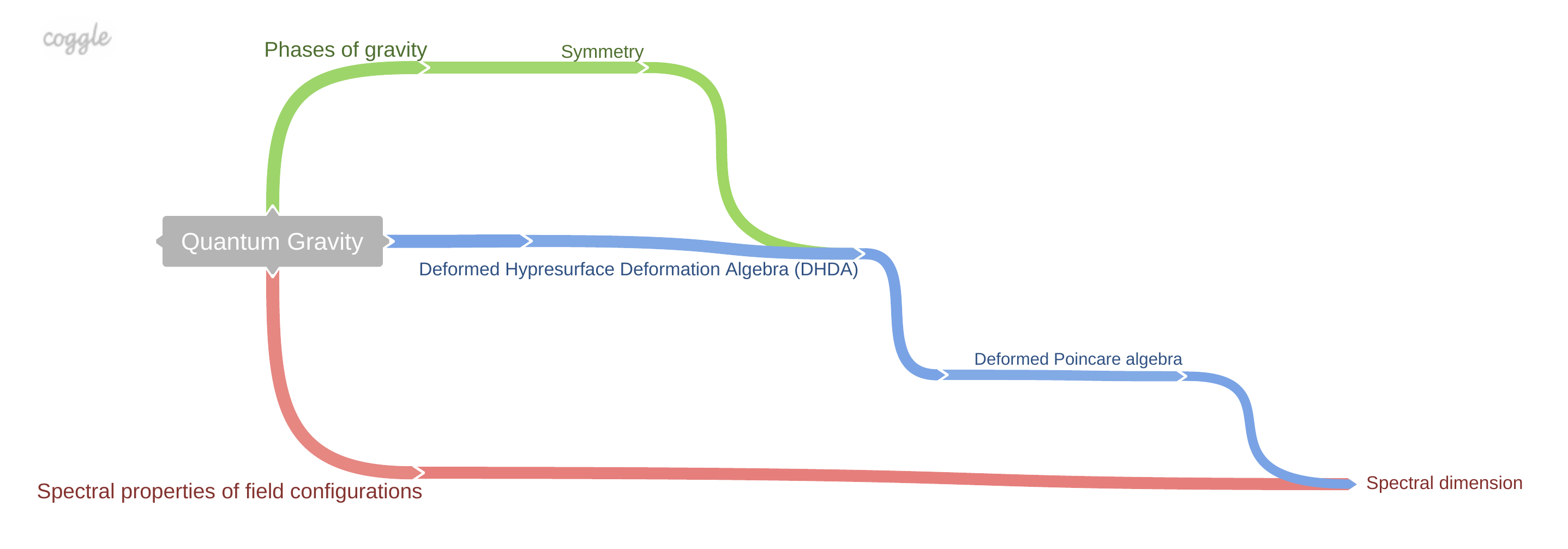}
\end{center}
\caption{Exemplary architecture of some aspects of quantum gravity.}
\label{Fig2}
\end{figure}

In Fig.~\ref{Fig2} we present a piece of the architecture that we are going to study here. 
The initial node in our map is ``quantum gravity''. From there one can pass to the concept of 
``phases of gravity''. Indeed, gravity is a system with a huge (perhaps infinite) 
number of interacting degrees of freedom. Systems of this kind, occurring 
in the physics of complex systems, generically exhibit such emergent phenomena 
as the \emph{nontrivial phase structure}. Various phases can arise, 
depending on the surrounding environment. The particular feature distinguishing 
different phases is a ``symmetry'' -- our second building block. An 
analogy that can be recalled here is the difference between the solid and liquid states 
of water. The symmetries of the liquid state are the invariance under rotations and translations, 
while in the solid state these symmetries are partially broken due to the formation of the crystal structure. 

The notion of symmetry can be expressed in the algebraic terms by introducing 
the generators of symmetry transformations. For example, the generators of rotations form the 
$\mathfrak{so}(3)$ algebra. This leads us to the third block of our construction, namely the 
``Deformed Hypersurface Deformation Algebra'' (DHDA). Let us explain how does it arise. 
Classical theory of gravity -- General Relativity -- tells us that physics 
does not depend on the choice of a coordinate system. Such a property is a symmetry 
of the theory, which is known as general covariance. The generators of this symmetry 
(the scalar and vector constraints) form the first class algebra, the so-called
Hypersurface Deformation Algebra (HDA). On the other hand, when quantum gravity 
effects are taken into account, the structure of HDA may become deformed, so that it is 
replaced by a certain DHDA. 

HDA encodes the symmetry of any pseudo-Riemannian manifold but each manifold 
locally reduces to Minkowski spacetime, whose symmetries are described by the 
Poincar\'{e} algebra. The Poincar\'{e} algebra is, therefore, a sub-structure of HDA. 
This generalizes to the case of an arbitrary DHDA, which is expected to reduce to the 
corresponding deformation of the Poincar\'{e} algebra. Various versions of such a 
quantum deformed Poincar\'{e} algebra have been considered in the context of quantum gravity. 
It shows how the small scale structure of spacetime is modified by the quantum 
gravity effects. Another simple way in which these effects could manifest is a change 
in the effective dimensionality of spacetime. The latter phenomenon has been studied in 
different approaches to quantum gravity by investigating the variability (as a function of scale) 
of the dimension of spacetime, usually defined as the Hausdorff or spectral dimension. 
We will focus here on the latter, which quantifies the spectral properties of a given field 
configuration and explicitly depends on the form of the (deformed) Poincar\'{e} algebra 
and the associated Laplace operator. In the mathematical sense, the spectral dimension 
belongs to the ``spectral properties of field configurations''. 

In the next sections we will gather the ingredients necessary to built a more detailed 
realization of the architecture introduced above, which will be presented in Sec.~\ref{finalmap}. 
It should be stressed that in this realization we restrict to a particular selection of approaches 
from Fig.~\ref{Fig1}, especially the ones that we are most familiar with. The obtained 
construction will be hopefully extended in the sequel to this paper, where we would like 
to include other aspects of quantum gravity. Here we focus on understanding the 
relationships between: phases of gravity, the running dimension of spacetime, deformations 
of relativistic symmetries and the nonlinear structure of phase space.

\section{Phases of gravity} \label{sec:phases}

There is growing evidence (see e.g. \cite{Smolin:1996ca,Stephens:2001sd,Bojowald:2015qw,Mielczarek:2014aka}) that the 
gravitational field can exist in different phases (i.e. macroscopic states). This is actually not surprising, 
since fields as well as systems of particles usually form various phases, invariant under different sets of 
symmetries and determined by the initial conditions and interaction with the environment. The so-called analogue gravity models \cite{Barcelo:2005ay} have shown many interesting similarities between gravity and condensed matter physics. Furthermore, it is becoming 
apparent that the quantum gravitational phenomena might also be explained in terms of the quantum many-body systems, which naturally 
reflects their non-trivial phase properties \cite{Oriti:2017twl}. In the case of gravity, the first indication of the 
nontrivial phase structure came from the numerical results of Euclidean Dynamical Triangulations (EDT), 
formulated within the path integral quantization approach. EDT facilitate the computations 
by discretizing spacetime into simplices, with the help of the Regge calculus, and making the Wick 
rotation to the Euclidean domain. Consequently, the quantum gravitational field can be described by 
a statistical ensemble and studied using the Monte Carlo simulations, which allow to find the equilibrium 
configurations for different values of the coupling constants. An equilibrium is equivalent to a classical path, 
while thermal fluctuations correspond to quantum fluctuations around it. 

The (2+1)-dimensional EDT were the first to predict two distinct phases of gravity, 
separated by a first order phase transition \cite{Ambjorn:1991rx}: 
the so-called \emph{branched polymer} phase and the \emph{crumpled} phase. 
The investigations were later extended to 3+1 dimensions, showing that the above 
phase structure is preserved in this case \cite{Bialas:1996wu}. Gravitational phases can 
usually be distinguished by the effective dimensionality of spacetime (which will be 
discussed in more detail in Sec.~\ref{sec:dimension}). In particular, in the crumpled 
phase of EDT the Hausdorff dimension behaves as $d_H \rightarrow \infty$, 
while in the branched polymer phase we have $d_H = 2$ and the (constant) spectral 
dimension $d_S = \frac{4}{3}$. EDT in the standard formulation does not exhibit a phase with 
``extended'' four dimensional spacetime, which would describe a semiclassical solution 
of the theory. It has recently been verified \cite{Coumbe:2015em} that such a phase does not emerge even after introducing a nontrivial measure in the path integral. 

Further information about the presumed phase structure of gravity comes from simulations 
of Causal Dynamical Triangulations (CDT) \cite{Ambjorn:2005qm,Ambjorn:2012jv,Ambjorn:2018ty}. 
This improved approach is constructed by imposing a causal structure on 
configurations of EDT. The causality condition is realized owing to the introduction of a preferred 
foliation of spacetime, which may seem to be somewhat restrictive. However, it has also been 
argued \cite{Jordan:2013cn} that the preferred foliation is not a necessary ingredient but only a 
convenient choice and the results of CDT are indeed sufficiently generic. One of the remarkable 
consequences of CDT is appearance of the ``extended'' four dimensional phase. 
In total, three main phases, called $A$, $B$ and $C$, have been observed. An average 
spacetime configuration in the phase $C$ is a well extended blob characterized by the Hausdorff 
dimension $d_H \approx 4$. Moreover, most of the vertices of such a triangulation 
have relatively small valence, while the maximal valence in a spatial slice grows proportionally to its volume \cite{Ambjorn:2017ns}. The notion of geometry emerges naturally via a global coordinate system that 
can be introduced to parametrize the graph representing a given triangulation. The graph corresponding to the phase $B$ is lacking the latter 
property since in this case the typical valence of vertices is too large. According to the numerical results of 4d 
CDT, this phase is characterized by the Hausdorff and spectral dimensions tending to infinity, 
$d_H \rightarrow \infty$, $d_S \rightarrow \infty$. Therefore, the phase $B$ is often perceived 
as a counterpart of the crumpled phase in 4d EDT. On the other hand, the phase $A$ shares 
some properties of the branched polymer phase. However, a detailed study of this phase has 
not been performed so far. 

The phases $C$ and $A$ are separated by a first order transition line \cite{Ambjorn:2012ij}, 
while the transition between $C$ and $B$ is of the second order \cite{Ambjorn:2011cg}. The order 
of the $A-B$ transition is unknown. The $C-A$, $C-B$ and $A-B$ transition lines meet at the 
\emph{triple point}. Furthermore, recent results in 4d CDT suggest that there exists an additional phase 
inside the phase $C$, called the \emph{bifurcation} phase \cite{Ambjorn:2017ns}, which seems to 
exhibit the phenomenon of the metric signature change \cite{Ambjorn:2015qja}. Further analysis is required 
to understand the nature of this new region on the phase diagram.      

It should also be stressed here that different types of the order parameter can be applied in order to discern the 
phases and transitions between them. Each order parameter is sensitive to certain particular features (including symmetries) 
of a given field configuration. The $A,B,C$ phases in CDT are detected using the heuristically introduced 
``average geometry" parameter. However, in order to observe the bifurcation phase, which is 
a subphase of the phase $C$, a more sophisticated parameter has to be chosen. 
Therefore, the number of perceived distinct phases depends on what type of the order parameter is considered. 

Interestingly, it has been shown \cite{Wilkinson:2014zga,Wilkinson:2015fja} that three phases 
similar to those present in 4d CDT appear also in the so-called Quantum Graphity \cite{Konopka:2008hp} 
model of the Planck scale physics. In this case the (extended/semiclassical) phase $C$ is realized at the 
minimum of energy and is the most stable configuration (i.e. a vacuum). The phases $A$ and $B$ can be 
reached by departing from the minimal energy. Furthermore, the transition between 
the high temperature non-geometric phase and low temperature geometric phase shares 
properties of the transition between the crumpled phase $B$ and the geometric phase $C$ in CDT. 
In this context the term \emph{geometrogenesis} has been coined and possible observational 
relevance of such gravitational phase transitions has been studied \cite{Magueijo:2006fu}. 
One can say that the $B-C$ transition in CDT provides a concrete realization of the 
geometrogenesis discussed in the context of Quantum Graphity. Cf. a recent study
\cite{Mandrysz:2018us}, which also explores the relation between geometrogenesis and the ultralocal limit of gravity (we will discuss the latter in Sec.~\ref{sec:dimension}). 

Transitions between different phases of gravity in principle may be of the first, second or higher order. 
Among them, the second order transitions deserve a special attention. The reason is that at 
such a transition the correlation function of a given order parameter diverges and the 
theory becomes scale invariant. Consequently, field configurations in the continuum limit 
(of discretization) should be described by a conformal quantum field theory. This concerns either 
the full spacetime geometry or only its spatial part. 

Therefore, it is clear that GR, which does not satisfy the conformal invariance, cannot 
describe the classical state of gravitational field at a second order transition point (or line, as in the case of 
CDT). We may suppose that GR is just a (semi)classical theory of the extended phase of gravity, 
while at the phase transitions, or in other phases, the fundamental quantum theory of gravity 
reduces to certain effective theories that are different from GR. Let us compare this possibility with the 
phase structure of water. The fundamental quantum Hamiltonian of water is the same for every phase but 
various effective descriptions are needed in individual phases. In particular, the Clapeyron equation 
provides a sufficient effective model in the low-density gaseous state. On the other hand, the liquid 
state is well modeled as a non-compressive fluid, characterized by the Navier-Stokes equation. 
Both equations may be called the equations of state, as the relations between variables that are satisfied 
in a given phase. Analogously, perhaps the Einstein equation is an equation of state that is valid only 
in the extended phase of gravity. Then different equations of state would be required as the proper 
description of gravity in other regions of phase space. 

Coming back to the second order phase transition, a choice of the proper equation of state in such a case 
depends on whether it is full spacetime that is conformally invariant or only its spatial slices. For 
the full conformal invariance one might consider e.g. the Weyl gravity\footnote{The action 
of the Weyl gravity is $S = \frac{1}{16\pi G} \int d^4x \sqrt{-g} C_{\mu\nu\alpha\beta} C^{\mu\nu\alpha\beta}$,  
where $C_{\mu\nu\alpha\beta}$ denotes the Weyl tensor. The Weyl tensor is invariant under a 
conformal transformation $g_{\mu\nu} \rightarrow g'_{\mu\nu} \equiv \Omega^2(x) g_{\mu\nu}$.}. 
Therefore, a classical theory of gravity different than GR, such as the Weyl gravity (investigated so far due 
to the purely theoretical reasons), can potentially arise as an effective description of the gravitational 
field at the second order transition. On the other hand, when the conformal invariance holds only at the 
level of spatial hypersurfaces, the so-called Cotton tensor will be more adequate to consider. The 
latter situation occurs in the Ho\v{r}ava-Lifshitz approach to quantum gravity \cite{Horava:2009uw}. 
In this context it is worth to mention that, as analyzed in \cite{Ambjorn:2010hu}, the Ho\v{r}ava-Lifshitz 
theory and its precursor -- the Lifshitz scalar, have the phase structure analogous to the one observed 
in CDT for the ``average geometry" order parameter. 

The existence of various phases of gravity has been also predicted within Group Field 
Theory (GFT) \cite{Oriti:2014uga}, Loop Quantum Gravity and Spin Foam models. Worth mentioning 
here is that, although strictly related, these three approaches are characterized by different Hilbert 
spaces and dynamics. A GFT model defined on the group ${\rm U}(1)^{\times 3}$ was studied 
with the help of the Functional Renormalization Group and found to have the nontrivial phase structure, 
containing the Gaussian and non-Gaussian fixed points \cite{Benedetti:2014qsa}. Furthermore, 
the emergence of bosonic condensate phases of GFT, which are probably associated with the vacuum states different from the Fock vacuum, has been observed and interpreted 
in the cosmological terms, cf. \cite{Gielen:2016qw,Oriti:2016acw} and references therein. Such GFT condensates are another potential realization of the geometrogenesis mentioned above. Meanwhile, the phase diagram of Spin Foam models 
was tentatively derived in \cite{Delcamp:2016dqo}. An indication of distinct phases in LQG can also be noticed 
within investigations of its vacuum. In the recently proposed new formulation of LQG 
\cite{Dittrich:2014wpa} there appears a new vacuum state, the so-called BF-vacuum $|0\rangle_{BF}$. 
Such a ground state (given by a spin network with a large number of vertices) has the constant curvature 
and is a sort of the condensate state, similar to the ones considered in GFT. This contrasts with the 
Ashtekar-Lewandowski vacuum $|0\rangle_{AL}$, for which no single node of a spin network exist. 
The two vacuum states correspond to two inequivalent representations of the LQG algebra of canonical 
variables. Therefore, they are possibly associated with different phases of the gravitational field. 
Finally, the Random Tensor Models, which are the higher dimensional extension of the Random 
Matrix Models used to study 2D quantum gravity and can also be seen as a simple version of GFT, indicate the non-trivial phase 
structure via their observed critical behavior (in the large $N$ expansion) \cite{Bonzom:2011zz,Bonzom:2014ts}. Further discussion and examples 
of the many-body aspects of quantum gravity and the related non-trivial phase structure can be found 
in e.g. \cite{Bojowald:2015qw,Oriti:2017twl,Oriti:2018tym}.

\section{Asymptotic safety and UV fixed point}  

There is much more to say about the second order phase tradition in gravity. Namely,
this type of a transition is associated with existence of the critical point 
(or critical line). The criticality leads to the hypothesis of \emph{asymptotic safety} 
\cite{WeinbergAS,Percacci:2007sz}, first proposed by Weinberg (see below). 

General Relativity is perturbatively nonrenormalizable, which is caused by the dimensionful character of 
the Newton coupling constant. In a perturbative expansion of the probability amplitudes there occur 
additional UV divergences, which require a regularization by adding the appropriate counterterms to the 
Hamiltonian. Consequently, new coupling constants (that multiply counterterms) appear in the quantum 
theory and have to be fixed in order to make predictions. Since values of these couplings are a priori 
unknown, the theory is lacking the predictive power. 

While GR is perturbatively nonrenormalizable, there is still a possibility that it is nonperturbatively 
renormalizable. This is precisely the idea of asymptotic safety. Namely, the conjecture put forward by 
Weinberg \cite{WeinbergAS} is that if there exists a non-Gaussian UV fixed point in the Renormalization 
Group (RG) flow and it can be reached by trajectories belonging to a finite dimensional critical surface, 
then only a finite number of couplings (equal to the dimension of the critical surface) has to be 
fixed to cancel the divergences of GR. Investigation of such a scenario has attracted 
significant attention in recent years \cite{Reuter:2012id}. 

Analysis of various effective quantum actions of GR (containing higher order derivatives) 
supports the asymptotic safety conjecture by indicating existence of a critical point (nontrivial fixed 
point) for the considered actions. However, in order to prove the conjecture, all allowed 
effective actions should be taken into account. If the conjecture is correct, then there would be no formal 
necessity to introduce new degrees of freedom for gravity at the Planck scale and GR could be quantized 
just as it is. Furthermore, in the latter case there would be no interpretation of GR as an effective theory, 
valid only at the energy scales below the Planck energy, as one may expect from its perturbative 
nonrenormalizability. Nevertheless, it is important to stress that asymptotic 
safety does not rule out that GR actually is an effective theory. It is still possible that more 
fundamental degrees of freedom have to be introduced due to the underlying physical reasons. Such a situation happens 
in the theory of hydrodynamics, which despite being simultaneously perturbatively nonrenormalizable and 
asymptotically safe, does not provide the accurate description of liquid water at the atomic scale or near the phase transitions. 

Searching for the critical behavior of the gravitational field gains, therefore, an additional motivation. 
In fact, the CDT results discussed in the previous section seem to support the asymptotic 
safety conjecture. This is due to the presence of the second order transition line. 
Preliminary results of the RG analysis in CDT \cite{Ambjorn:2014gsa} indicate that the 
triple point, which ends the second order $C-B$ transition line, is probably related 
to the UV critical point via the continuum limit. However, then there may appear the anisotropic scaling 
between spatial and temporal directions, as suggested in \cite{Ambjorn:2014gsa}. 
In such a situation, RG trajectories would ``flow'' from approximately the centre of the phase $C$ 
towards the second order transition line and finally converge at the triple point. 
Further numerical simulations have to be performed in order to verify these findings. 

The type of anisotropy that is needed to obtain a UV fixed point in CDT is exactly the one 
assumed in the Ho\v{r}ava-Lifshitz gravity (HLG) approach to quantum gravity. HLG is a 
generalization of GR whose action (in the UV) is invariant under the anisotropic scaling of spacetime coordinates 
\cite{Horava:2009uw}
\begin{equation}
{\bf x} \rightarrow b\, {\bf x}\,, \qquad t \rightarrow b^z t\,,
\end{equation}
with an arbitrary constant $b$ and the critical exponent $z$. The standard GR can be recovered for $z = 1$, while for 
$z = 3$, at the so-called \emph{Lifshitz point} of the RG flow, 
the theory in 3+1 dimensions (similar to a Lifshitz scalar field theory) becomes power-counting renormalizable. This and other 
choices of the critical exponent will be discussed in more detail in the next section. 

An extremely promising possibility is that the UV fixed point appearing in the 
asymptotic safety scenario, the triple point on the CDT phase diagram and the 
Lifshitz point ($z = 3$) of HLG are actually the same critical point. 
Besides the arguments presented above, further support for such a case comes 
from the results for the spectral dimension of spacetime. 

\section{Spectral dimension and phases of gravity} \label{sec:dimension}

A method to characterize the structure of quantum spacetime that has been widely used 
in recent years are calculations of the effective number of spacetime dimensions, especially 
applying the notion of the spectral dimension. This particular definition of the dimension 
employs the fact that the return probability of diffusion (random walk) to the same point 
strongly depends on the dimensionality of a manifold on which it is considered. In particular, 
in Euclidean space $\mathbb{R}^d$ the averaged (over the whole space) return probability 
scales as $P(\sigma) \propto \sigma^{-d/2}$, where $\sigma$ is the auxilliary diffusion time. 
By analogy with the scaling of $P(\sigma)$ in Euclidean space, 
the \emph{spectral dimension} of a manifold is introduced as
\begin{equation}
d_S := -2 \frac{\partial \log P(\sigma)}{\partial \log\sigma}\,, 
\label{SpectralDim}
\end{equation}
so that for Euclidean space we have simply $d_S = d$. Since the above definition concerns Riemannian manifolds, spacetime first has to be Wick-rotated. Moreover, it should be mentioned that the expression (\ref{SpectralDim}) does not 
take into account the effects of topology or curvature. 

The scaling of $P(\sigma)$ is a probe of departure from the Euclidean geometry, which is 
associated with the spectral properties of a given manifold. In order to determine this quantity one considers a fictitious diffusion process, described by 
the heat 
equation
\begin{equation}
\frac{\partial}{\partial \sigma} K({\bf x},{\bf y}; \sigma) = \Delta_x K({\bf x},{\bf y}; \sigma)\,, 
\label{HKE}
\end{equation}
where $\Delta_x$ denotes the appropriate Laplace operator 
and the initial condition $K({\bf x},{\bf y}; \sigma) = \delta^{(d)}({\bf x - y})$ 
is assumed (when the manifold is flat). The heat kernel $K({\bf x},{\bf y}; \sigma)$, which is the solution to (\ref{HKE}), allows 
to calculate the average return probability
\begin{equation}
P(\sigma) := \int dx\, K({\bf x},{\bf x}; \sigma) = \int d\mu(p)\, e^{\sigma \Delta_p}\,, 
\label{returnprobability}
\end{equation}
where $\Delta_p$ is the momentum representation of $\Delta_x$ and $\mu(p)$ 
an invariant measure on momentum space. For quantum spacetimes 
it is required that at large diffusion times (probing large scales) the classical value of the dimension 
is recovered. On the other hand, for small times the small 
scale structure of spacetime is explored, corresponding to the UV limit of the theory, which may lead to the nontrivial behavior of the dimension. 

Indeed, running of the spectral dimension as a function of scale has been found in 
such approaches to quantum gravity and models of quantum spacetime as: CDT \cite{Ambjorn:2005db,Coumbe:2014noa}, 
Ho\v{r}ava-Lifshitz gravity \cite{Horava:2009if}, asymptotic safety scenario \cite{Lauscher:2005fy}, nonlocal quantum gravity \cite{Modesto:2012sy}, 
broadly understood loop quantum gravity \cite{Modesto:2009fm,Calcagni:2014cza} (so far, only at the kinematical level), 
causal sets \cite{Carlip:2016dy,Belenchia:2016ss}, multifractional 
spacetimes \cite{Calcagni:2012qn}, noncommutative spacetimes \cite{Benedetti:2009fe,Arzano:2014de,Arzano:2017ua} 
and spacetimes characterized by a deformed hypersurface deformation algebra \cite{Mielczarek:2016zfz} (see the next section). In almost all of the 
cases the obtained results show a dimensional reduction in the UV limit, usually to 
$d_S \approx 2$, which is the value that makes gravity power-counting renormalizable. This is also in agreement with what is expected at the UV fixed point in the asymptotic safety conjecture 
\cite{Lauscher:2005fy}. Other results may be associated with different phases of the theory. In general, spacetimes exhibiting the phenomenon of the running dimension are called multiscale. Various relevant aspects of the 
dimensionality of spacetime are discussed e.g. in \cite{Carlip:2017dy} and within a recent review of multifractional spacetimes \cite{Calcagni:2017mw}. 

\begin{figure}
\begin{center}
\includegraphics[width=16cm]{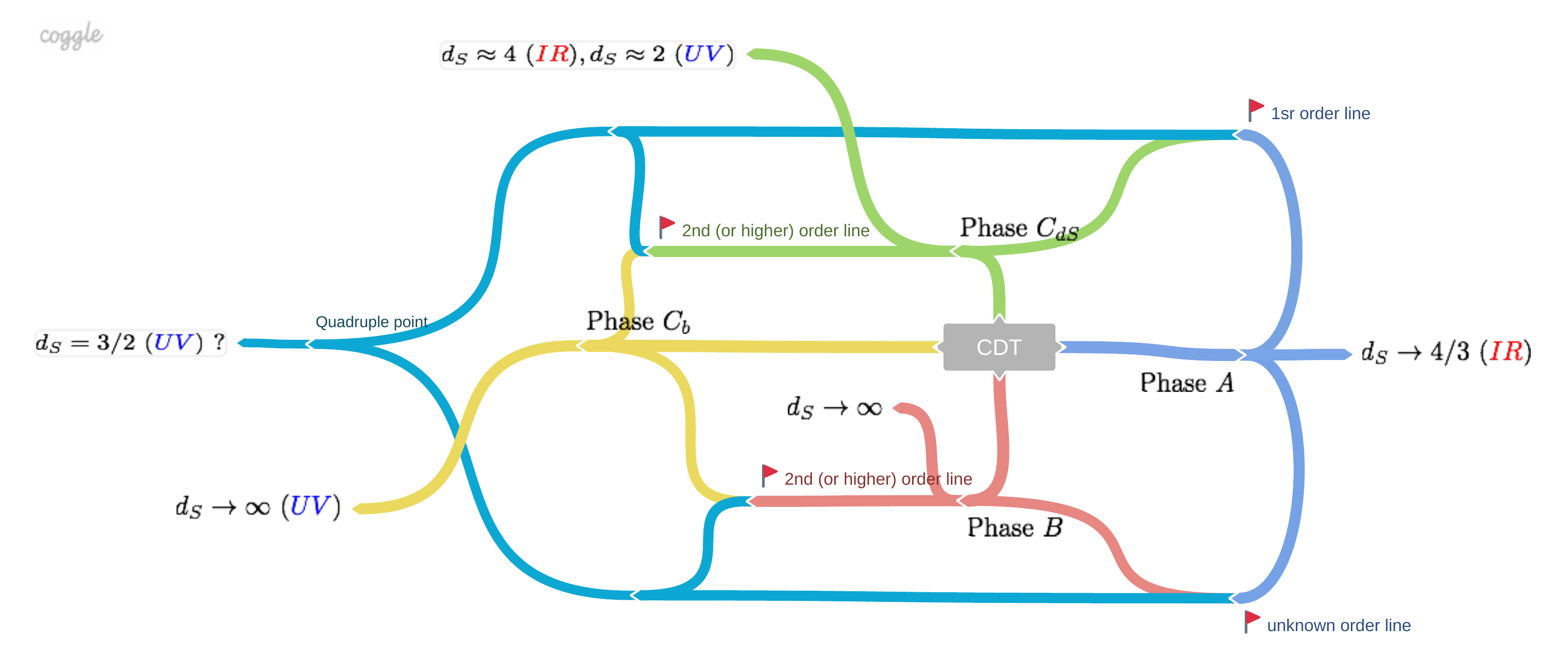}
\end{center}
\caption{Phase structure of CDT and asymptotics of the 
spectral dimension.}
\label{CDTPhasSpect}
\end{figure}

In the case of CDT, each of the four phases (taking into account the bifurcation phase) is 
characterized by a different scale dependence of the spectral dimension. 
All phases and transitions between them 
have been collected on a graph in Fig.~\ref{CDTPhasSpect} together with the corresponding IR and UV limits of the spectral dimension. 

As it was already mentioned in Sec.~\ref{sec:phases}, in the non-geometric phase $B$ the UV 
value of the spectral dimension tends to infinity. This is a consequence of the extremely high 
connectivity between different points of space. It takes only a few Planck steps to go from one 
point to any other since for small diffusion times (for small number of steps) there is always a 
huge number of neighboring points. Spacetime is effectively very high dimensional, which is reflected 
in the spectral dimension. On the other hand, it is also easy to return in several steps to the 
starting point. Therefore, while the spectral dimension is sharply peaked at small diffusion times, 
it quickly falls to zero for larger times, as it was indicated by simulations performed in CDT. 
This shows that the crumpled spacetime configuration in the IR looks like a single space-time point. 

Spacetime in the extended phase $C$ has a 
completely different structure, which can be interpreted as semiclassical. At large scales, i.e. in the IR limit, the 
spectral dimension saturates at $d_S \approx 4$, recovering (when a contribution of the assumed compact 
topology is subtracted) the value from the classical theory. In the seminal paper \cite{Ambjorn:2005db} it 
was also discovered that the dimension monotonically decreases with scale and small scales (the UV limit) are 
characterized by the value $d_S \approx 2$. Based on the requirement of consistency, one may speculate 
that the same behavior occurs at the triple point on the phase diagram but this presumption has to be verified 
by further numerical studies. Furthermore, as more extensive simulations have recently shown \cite{Coumbe:2014noa}, 
the UV limit of the spectral dimension actually depends on a location within the phase $C$. In particular, it has 
been found that the value $d_S \approx 2$ is being measured in a region lying deep inside the phase $C$. 
However, if one approaches the transition line to the phase $A$, the spectral dimension decreases to 
$d_S \approx \frac{3}{2}$, which probably reflects the renormalization group flow. 

Finally, the spectral dimension in the phase $A$ has not yet been a subject of systematic 
studies. In the numerical simulations spacetime in this phase appears to behave as a sequence of causally 
disconnected spatial slices. A preliminary analysis of the spectral dimension indicates that 
at large scales it becomes $d_S \rightarrow \frac{4}{3}$ \cite{Coumbe:2014noa}, which is the same 
as for the \emph{branched polymer} phase in EDT. 

Meanwhile, in HLG (the Ho\v{r}ava-Lifshitz gravity) it has been found that the spectral 
dimension depends on the value of the (running) critical exponent $z$ and is given by \cite{Horava:2009if}
\begin{equation}
d_S = 1 + \frac{D}{z}\,, \label{dsHL}
\end{equation}
where $D$ denotes the topological dimension of space, which we fix here as $D = 3$ (i.e. we consider (3+1)-dimensional spacetime). It is clear 
that the classical case with $d_S = 4$ is correctly recovered for $z \rightarrow 1$. In turn, 
at the Lifshitz point $z \rightarrow 3$ the spectral dimension reduces to $d_S = 2$. 
The latter result coincides with the UV 
limit of $d_S(\sigma)$ in the central region of the phase $C$ in CDT. Indeed, at least in the case of 2+1 dimensions, it has been shown 
to high accuracy \cite{Sotiriou:2011ss} that HLG can reproduce the spectral dimension in the above region on the CDT phase diagram 
for the range of scales lying between the UV and IR (although this is not enough to prove that the small scale structure of spacetime in 
the two theories is identical \cite{Calcagni:2013pn}). Moreover, the case of $d_S = \frac{3}{2}$, which occurs in 3+1 dimensions in both 
CDT and EDT, in the Ho\v{r}ava-Lifshitz gravity is obtained for $z = 6$. Analysis of the theory with this value of $z$ has not been carried 
out so far. Furthermore, in Ho\v{r}ava-Lifshitz gravity the value of spectral dimension corresponding to the branched polymer case (or CDT 
phase A) is obtained for $z = 9$.

There are two more special situations in HLG that should be explored, namely $z \rightarrow 0$ and $z \rightarrow \infty$. In particular, 
analysis of the scaling properties of the gravitational action leads to the conclusion that the case $z \rightarrow 0$ realizes the so-called 
\emph{ultralocal limit} \cite{Isham:1976} of gravity (see also Sec.~\ref{sec:DHDA}), in which spacetime splits into a congruence of causally 
independent worldlines. This is a consequence of suppressing the spatial derivatives in the action, so that only the kinetic and cosmological 
constant terms remain. According to the formula (\ref{dsHL}), the value of the spectral dimension in such a limit will tend to infinity, 
$d_S \rightarrow \infty$. The result may seem counterintuitive since in the ultralocal state of spacetime the spectral dimension is expected 
to reduce to $d_S = 1$, corresponding to the remaining single direction. The possible explanation is found by considering the phase $B$ of 
CDT, which can be seen as a particular realization of the ultralocal state, characterized by the anisotropic scaling $z \rightarrow 0$ (see the 
next paragraph). As we already discussed, in the phase $B$ the dimension at small scales becomes $d_S \rightarrow \infty$ but at sufficiently 
large scales spacetime is effectively a point. 

Comparing the ultralocal state and the phase $B$ of CDT one can suspect that they are closely related. Indeed, it has been shown 
\cite{Ambjorn:2010hu} that when the phase $B$ is approached from the phase $C$, the ratio of lengths of spacelike and timelike simplicial 
links is decreasing. Such an effect is associated with the collapse of lightcones into worldlines, which is exactly what happens in the ultralocal 
limit. The ultralocality of the phase $B$ is, however, partially obscured due to the constraints that are imposed at the level of numerical simulations. 
The crucial assumption of CDT is that the (compact) topology of spatial slices is preserved, which is necessary to ensure that the causality in 
quantum spacetime is not violated. However, this constraint also prevents the gravitational configuration from achieving the ``extended'' ultralocal state, in 
which all points of space evolve as disconnected universes. Another restriction, introduced for computational reasons, is that the total number 
of simplices (or the average of it) is fixed, being controlled by the cosmological constant, which plays the role of a chemical potential in the statistical ensemble. Due to 
the combined effect of the above constraints in CDT, an attempted simulation of the ultralocal state leads to a pointlike universe pierced by a timelike 
straw (the latter may or may not have the physical meaning). The dimensionality of space in the IR is correctly zero, which 
results from the crumpled structure in the UV. The only possible real difference with respect to the ultralocal state in HLG is the behavior of the time direction. Depending on whether 
the straw observed in numerical simulations is interpreted as a physical feature or a numerical artifact, the time direction 
stretches a single dimension or disappears, respectively. Let us also mention that since the phase $B$ 
corresponds to the anisotropic scaling $z = 0$, while the phase $C_{dS}$ to $z = 1$, we conjecture that the (sub-)phase $C_b$ should correspond to the 
scaling that interpolates between these two values of $z$. However, the properties of the phase $C_b$ have not yet been completely studied. 

Coming back to the Ho\v{r}ava-Lifshitz gravity, the last case to consider is the critical exponent $z \rightarrow \infty$, 
for which the spectral dimension $d_S \rightarrow 1$. While the dimension behaves here as it can be expected for 
the ultralocal state, field configurations are not characterized by vanishing of the spatial derivatives. It is quite the 
opposite, since the action invariant under the $z \rightarrow \infty$ scaling contains derivatives of the infinitely high 
order, possibly describing an extremely strongly correlated configuration. Physical properties of the case $z \rightarrow \infty$ 
deserve a separate detailed analysis, which still has to be carried out. 

\begin{figure}
\begin{center}
\includegraphics[width=16cm]{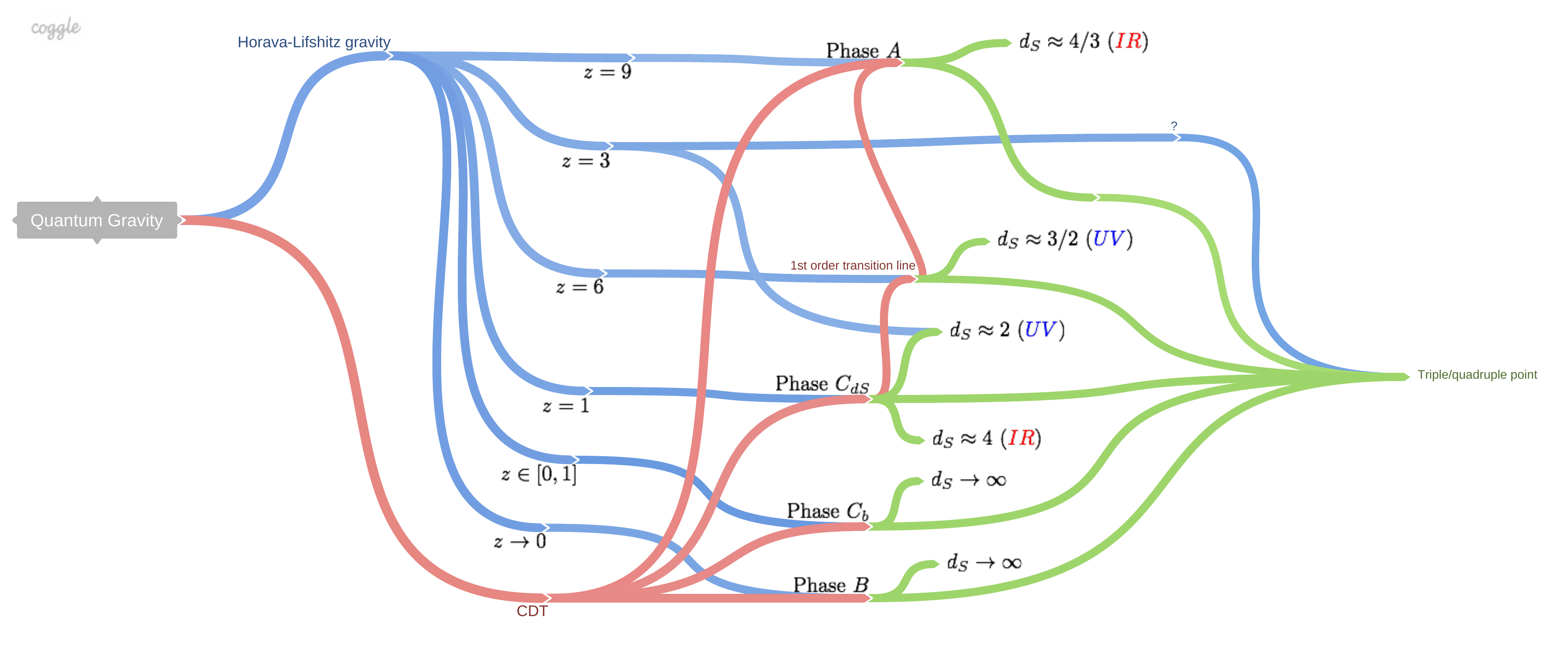}
\end{center}
\caption{Relations connecting CDT and the Ho\v{r}ava-Lifshitz gravity.}
\label{CDTHL}
\end{figure}

Concluding the discussion, we can tentatively consider that the Ho\v{r}ava-Lifshitz gravity and Causal Dynamical Triangulations 
are connected by the network of relations depicted in Fig.~\ref{CDTHL}. 
In this section we actually restricted mainly to the above two approaches to quantum gravity. An outline of the more inclusive map of quantum 
gravity that is based on predictions for the dimensionality of spacetime (using also other notions besides the spectral dimension) has 
recently been proposed in \cite{Calcagni:2016as}.

\section{Deformations of the Hypersurface Deformation Algebra} \label{sec:DHDA}

The symmetry of General Relativity is general covariance. In the Hamiltonian framework 
it is embedded in the structure of the algebra of constraints, the so-called 
\emph{Hypersurface Deformation Algebra} (HDA). This algebra (more precisely, it is an algebroid \cite{Bojowald:2016hgh}) 
is of the first class and therefore the constraints are given the interpretation of symmetry generators.  In the case 
of classical GR transformations generated by the constraints coincide with the Lie 
derivatives. However, the evidence gathered in recent years suggests that this property 
may be spoiled when the Planck scale physics is taken into account. 

Namely, the effects of quantum gravity can lead to certain deformations (i.e. modifications of the brackets) of the hypersurface 
deformation algebra. The ensuing \emph{Deformed Hypersurface Deformation Algebra} (DHDA) 
remains of the first class and the constraints still act as the generators of gauge 
transformations. Nevertheless, apart from certain limits, the transformations will differ from the 
Lie derivatives. Consequently, the spacetime metric will no longer be a covariant object, which suggests that it 
does not play such a significant role here as in the classical theory. The deformations of this 
type have been derived through an analysis of the effective algebra of quantum 
constraints in loop quantum gravity. For both spherically symmetric configurations and 
cosmological perturbations the obtained algebra has the brackets \cite{Bojowald:2011aa,Bojowald:2012ux}
\begin{eqnarray}
\left\{D[N^a_1], D[N^a_2]\right\} &=& D[N_1^b \partial_b N^a_2 - N_2^b \partial_b N^a_1]\,, \nonumber\\
\left\{S^Q[N], D[N^a]\right\} &=& -S^Q[ N^b \partial_b N]\,, \nonumber\\
\left\{S^Q[N_1], S^Q[N_2]\right\} &=& D \left[s \Omega\, g^{ab}(N_1 \partial_b N_2 - N_2 \partial_b N_1)\right], 
\label{DHDAbra}
\end{eqnarray}
where $\Omega$ is the deformation factor (affecting only the last bracket), $g^{ab}$ denotes the spatial metric and 
$s$ the spacetime metric signature, which is $s = 1$ in the Lorentzian case and $s = -1$ in the Euclidean one. 
The superscript $Q$ means that the scalar constraints $S^Q[N]$ are themselves quantum deformed 
(with respect to their classical counterparts). In the case of cosmological perturbations it has been found that the 
deformation factor takes the following form \cite{Cailleteau:2011kr}
\begin{equation}
\Omega = \cos(2\gamma \bar{\mu} \bar{k}) = 1 - 2 \frac{\rho}{\rho_c} \in [-1,1]\,, 
\label{OmegaLQC}
\end{equation}
with the critical energy density $\rho_c = 3/(8\pi G \gamma \Delta) \sim \rho_{\rm Pl}$, depending on the 
Immirzi parameter $\gamma$ and the minimal surface area $\Delta$. 
Similarly, for the spherically symmetric configurations we have $\Omega = \cos(2\delta K_\varphi)$. In both 
cases the expression for $\Omega$ is a cosine function of the extrinsic curvature. Such an effect 
originates from the so-called holonomy corrections, which are the result of replacing the Ashtekar 
connection by the corresponding holonomies\footnote{The holonomy of the Ashtekar connection along a 
contour ${\cal C}$ is given by the path-ordered exponential $h_{\cal C} := {\cal P} e^{\int_{\cal C} A}$.}. 
However, the inverse volume corrections are to be expected as well, which will modify the 
expression (\ref{OmegaLQC}) \cite{Cailleteau:2013kqa}. It has also been shown 
\cite{Bojowald:2014zla} that substituting a function $f(p)$ in the place of the kinetic term 
$p^2$ in the field Hamiltonian leads to the DHDA with the deformation 
factor
\begin{equation}
\Omega = \frac{1}{2} \frac{d^2f(p)}{dp^2}\,.
\end{equation}
In particular, when the polymer quantization is applied to the canonical momentum via the function
\begin{equation}
f(p) = \frac{\sin^2(\lambda p)}{\lambda^2}
\end{equation}
(as it is done in loop quantum cosmology), one obtains $\Omega = \cos(2\lambda p)$. $\lambda$ is here 
the scale of polymerization, introduced in such a way that the classical symmetries are recovered for $\lambda \rightarrow 0$. This explains 
the origin of the cosine form of deformations observed in the above mentioned effective models of LQG. 

\begin{figure}
\begin{center}
\includegraphics[width=16cm]{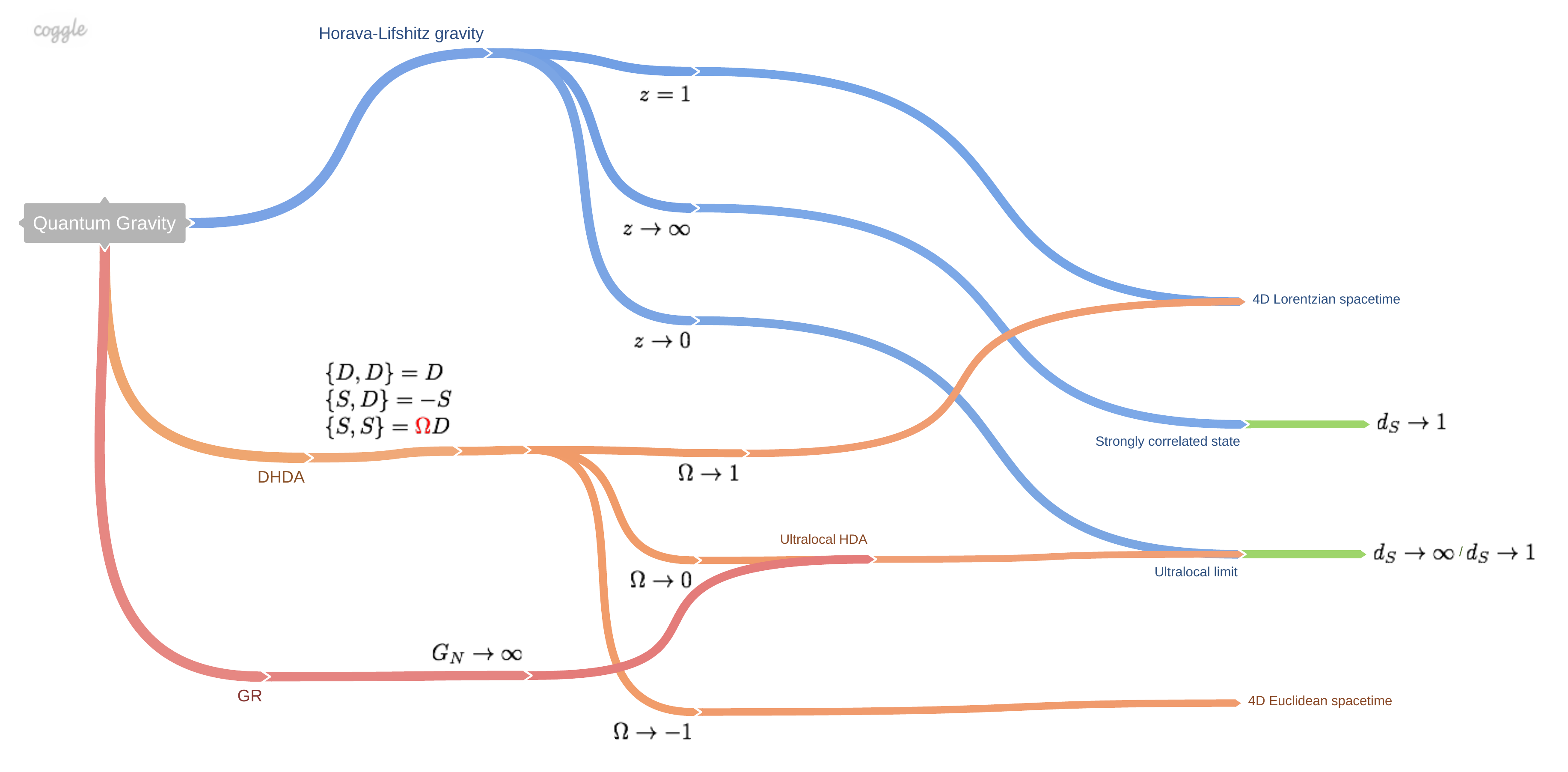}
\end{center}
\caption{Deformed HDA in the context of quantum gravity.}
\label{DHDAgraph}
\end{figure}

The graph presented in Fig.~\ref{DHDAgraph} contains different cases of the DHDA (\ref{DHDAbra}) and their relations with some other aspects of quantum gravity. 
Depending on the value of $\Omega$, several physical scenarios are realized. 
In particular, for $\Omega \rightarrow 1$ we recover classical GR with the Lorentzian metric signature. For $\Omega \rightarrow -1$ classical GR is recovered as well but 
with the Euclidean signature, while the case of $\Omega \rightarrow 0$ is equivalent to the 
ultralocal limit, discussed in the previous section. It is worth to add here that the ultralocal state can also be obtained 
in classical GR, by taking the strong coupling limit $G \rightarrow \infty$ \cite{Isham:1976}. 
Furthermore, in the sense of the BKL conjecture \cite{Belinsky:1970ew,Belinsky:1982pk}, the 
ultralocality is a general prediction when evolution of the gravitational field towards a singularity is considered. In this context 
it is associated with the concept of \emph{asymptotic silence} \cite{Andersson:2004wp}. 

As can be seen from (\ref{OmegaLQC}), the type of DHDA obtained in loop quantum cosmology (LQC) 
leads to the effect of the metric signature change. When energy density of the matter content of universe reaches 
the value $\rho = \rho_c/2$, then $\Omega$ changes its sign from positive to negative, which has the 
interpretation of going from the Lorentzian to Euclidean signature. This curious phenomenon has been a 
subject of several investigations in recent years \cite{Mielczarek:2012pf,Bojowald:2015gra}. Furthermore, the 
ultralocal state that emerges at $\rho = \rho_c/2$ has been studied in \cite{Mielczarek:2012tn}. An essential 
feature of the $\Omega$-deformation in LQC, pointed out in \cite{Bojowald:2016hgh}, is that it is actually 
associated with the non-Riemannian geometry of spacetime. Further understanding on this small scale structure 
in the regime where deformations are significant may be achieved through an analysis of the corresponding 
deformations of the Poincar\'{e} algebra. This issue is the topic of the next section.

\section{Connection between DHDA and the deformed Poincar\'{e} algebra}

The hypersurface deformation algebra (HDA) describes symmetries of an arbitrary 
pseudo-Riemannian manifold and the simplest case of such a manifold is naturally the Minkowski 
spacetime. The isometries of Minkowski spacetime form the Poincar\'{e} algebra, 
which can also be obtained in the limit of linear hypersurface deformations in the corresponding HDA. 
Accordingly, any modification of the standard HDA is expected to affect the structure of 
the Poincar\'{e} algebra. In particular, this is predicted to be the case in LQG, where 
the effective algebra of constraints (which is equivalent to HDA) is deformed, as we have already discussed. 

It has been argued that the deformed Poincar\'{e} algebra corresponding to the DHDA (\ref{DHDAbra}) 
has the following form \cite{Mielczarek:2016zfz}:
\begin{eqnarray}
\left\{J_a,J_b \right\} &=& \epsilon_{abc} J^c\,, \label{DP1}\\
\left\{J_a,K_b \right\} &=& \epsilon_{abc} K^c\,, \\
\left\{K_a,K_b \right\} &=& -s_{eff} \epsilon_{abc} J^c\,, \label{DP3}\\
\left\{J_a,P_b \right\} &=& \epsilon_{abc} P^c\,, \\
\left\{J_a,P_0 \right\} &=& 0\,, \\
\left\{K_a,P_b \right\} &=& \delta_{ab} P_0\,, \\
\left\{K_a,P_0 \right\} &=& s_{eff} P_a\,, \label{DP7}\\
\left\{P_a,P_b \right\} &=& 0\,, \\
\left\{P_a,P_0 \right\} &=& 0\,, \label{DP9}
\end{eqnarray}
with the deformed signature factor $s_{eff} \equiv s\, \tilde{\Omega}$, which is a function of the algebra generators and becomes the metric signature $s_{eff} = s$ in the limits $\tilde{\Omega} \rightarrow \pm 1$. So far, the form of 
$\tilde{\Omega}$ has not been determined in general but some attempts have been 
made. In particular, by requiring the algebra to satisfy the Jacobi identities and $\tilde{\Omega}$ to be a separable function of $P_0^2$ and $P_i^2$ (which is a convenient Ansatz), one obtains the 
following expression \cite{Mielczarek:2016zfz,Mielczarek:2013rva}:
\begin{equation}
\tilde{\Omega} = \frac{P^2_0 - \alpha}{P^2_i - \alpha}\,, 
\end{equation}
where $\alpha \in \mathbb{R}$ is a free parameter associated with the energy scale of 
the deformation. In such a case the Euclidean mass Casimir element (with the proper 
classical limit) can be constructed as
\begin{equation}
{\cal C}^E = \frac{P_0^2 + {\bf P}^2}{1 - \alpha^{-1} {\bf P}^2}
\end{equation}
and the simplest possible guess concerning the form of the d'Alambert operator 
on momentum space is $\Delta_p := -{\cal C}^E$. Substituting the latter into (\ref{returnprobability}), we 
are able to calculate the spectral dimension of spacetime equipped with the deformed symmetry algebra (\ref{DP1}-\ref{DP9}). It has been found 
\cite{Mielczarek:2016zfz} that the dimension reduces from $d_S = 4$ 
at large scales to $d_S = 1$ at short scales, reflecting the 
ultralocality arising in the limit $\tilde{\Omega} \rightarrow 0$. 

\begin{figure}
\begin{center}
\includegraphics[width=16cm]{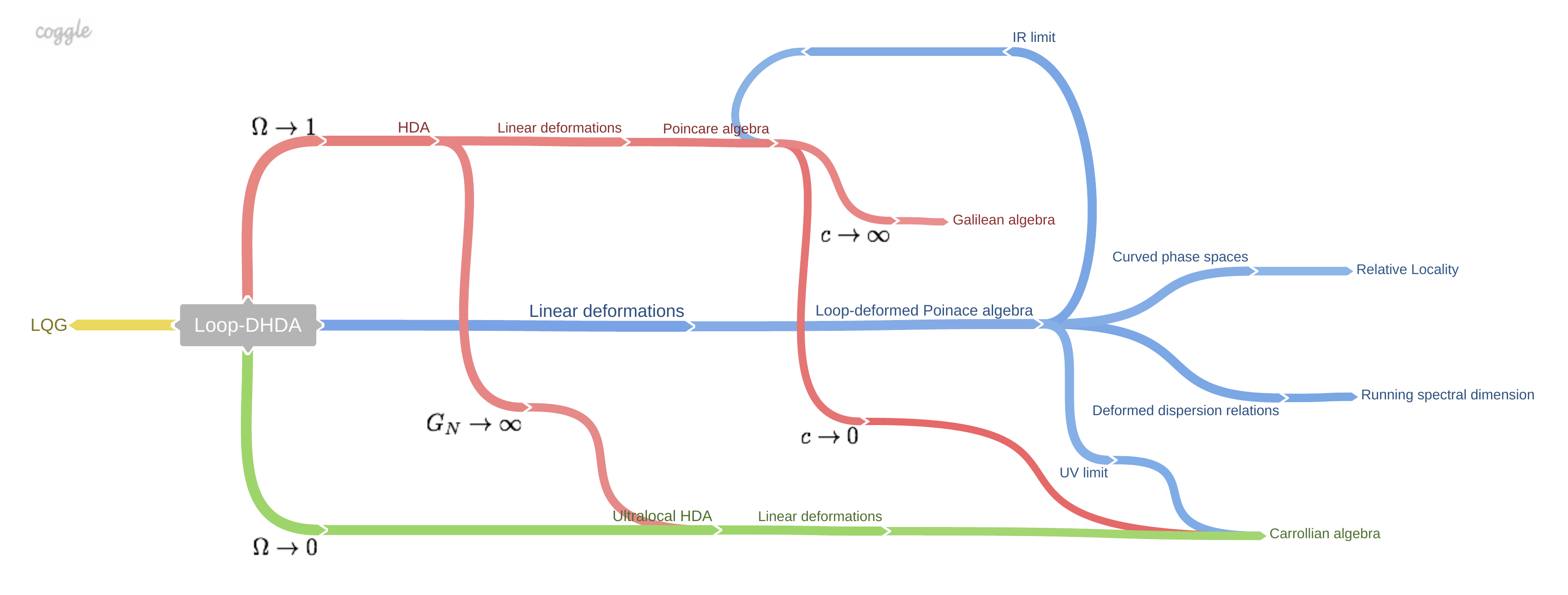}
\end{center}
\caption{From the DHDA to the deformed Poincar\'{e} algebra.}
\label{HDAPoincareBridge}
\end{figure}

The connection between HDA and the Poincar\'{e} algebra is a crucial element in 
the construction of our map of quantum gravity. When generalized to the deformed 
case, it allows to relate (see the end of this section) the two classes of, previously separated, approaches to the Planck 
scale physics: LQG and models of spacetime with deformed relativistic symmetries. This is illustrated in Fig.~\ref{HDAPoincareBridge}. 

In particular, let us focus on the DHDA given by (\ref{DHDAbra}). In the 
$\Omega \rightarrow 0$ limit it becomes an ultralocal algebra. By the restriction 
to the case of linear hypersurface deformations, this algebra reduces to the so-called Carroll algebra. 
So far, limited amount of research has been directed toward this structure. However, as we saw, 
it may actually reflect the quantum gravitational effects. The Carroll algebra has actually been introduced at the classical level, where it is 
obtained by taking the limit of vanishing speed of light, $c \rightarrow 0$, of the Poincar\'{e} algebra 
\cite{Levy-Leblond:1965}. This qualitatively overlaps with the observation made for 
DHDA, where the effective speed of propagation is found to be $v_{eff} = \sqrt{\Omega}$ 
and vanish for $\Omega \rightarrow 0$ \cite{Mielczarek:2012tn}. 

The general form of DHDA defined by the brackets (\ref{DHDAbra}) may lead to various deformations of the Poincar\'{e} 
algebra. In particular, in \cite{Amelino-Camelia:2016gfx} it was argued that the DHDA that was previously discussed in 
\cite{Bojowald:2012ux} possibly corresponds to the famous $\kappa$-Poincar\'{e} algebra \cite{Lukierski:1991qa,Lukierski:1992ny}. 
The latter algebra constitutes the archetypal example of quantum deformations of spacetime symmetries, 
which naturally appear in the semiclassical frameworks known as Doubly Special Relativity \cite{Magueijo:2001cr} and 
Relative Locality \cite{Amelino:2011py} (see below). Moreover, the form of deformations of both the Poincar\'{e} algebra 
and the corresponding HDA has been derived for certain multifractional spacetimes \cite{Calcagni:2017ds}. The results 
in these cases turn out to differ from (\ref{DHDAbra}). Meanwhile, the reconstruction of the appropriate DHDA in Relative 
Locality is an open task. An important role will probably be played there by the nonlinear momentum space, or nonlinear 
phase space, which is associated with the deformed relativistic symmetries.

\section{Nonlinear structure of phase space}

Deformations of relativistic symmetries discussed in the previous section are 
closely related to another concept appearing in the context of quantum gravity: momentum space or, more generally, 
phase space with the nontrivial geometry or topology. The origins of this idea go back to M.~Born, who argued that 
momentum space and the space of particle positions, 
i.e. spacetime, in quantum physics are connected by the \emph{reciprocity symmetry}. Therefore, 
in the regime of full quantum gravity, curved spacetime should be complemented 
by momentum space with similarly nontrivial Riemannian geometry \cite{Born:1938ay}. 
Later it was observed \cite{Snyder:1947qe} that non-vanishing curvature of 
momentum space leads to the noncommutativity of spacetime coordinates. 
This intuition was subsequently confirmed in the mathematical formalism of 
quantum groups, i.e. nontrivial Hopf algebras, which give a natural description 
of the deformed relativistic symmetries \cite{Majid:1995fy}. The most interesting case 
of such deformed symmetries is the $\kappa$-Poincar\'{e} (Hopf) algebra 
\cite{Lukierski:1991qa,Lukierski:1992ny}. This algebra acts covariantly on 
the noncommutative $\kappa$-Minkowski space \cite{Majid:1994by}, while the 
momentum space corresponding to the latter is the ${\rm AN}(3)$ Lie group, 
which as a manifold is equivalent to half of de Sitter space \cite{Kowalski:2002dy,
Kowalski:2003de}. 

The above mathematical structures have been utilized \cite{Amelino:2002re,
Amelino:2002dy,Magueijo:2001cr,Girelli:2005dy} in the framework known as \emph{Doubly} 
(or deformed) \emph{Special Relativity}, which is the family of models that attempt to represent 
the semiclassical regime of quantum gravity characterized by the existence of two 
invariant scales, given by the speed of light and Planck mass. These two scales 
determine the geometry of flat spacetime and curved momentum space. Doubly 
special relativity has recently been recast and generalized into the so-called 
\emph{Relative Locality} approach \cite{Amelino:2011py,Amelino:2011re}, whose 
underlying principle is that (at sufficiently high energies) only the full phase space 
is an absolute physical entity, while its decomposition into spacetime and momentum 
space depends on the choice of an observer. More specifically, the structure of spacetime is inferred 
from the dynamics of particles in (curved) momentum space. This leads to the relativity 
of locality of events in spacetime, i.e. physical events which have the same spacetime 
coordinates in the frame of a certain observer are not coincidental according to 
observers that are distant from the former. If we want to extend the above concept 
to the generally covariant case, both spacetime and momentum space should be allowed to have 
the nontrivial geometry. In order to study such a scenario there have been attempts to construct the 
action principle \cite{Cianfrani:2014ge} or the Hamilton geometry \cite{Barcaroli:2015hs} for particles having the nontrivial phase space. On the more 
fundamental level, there have been proposed \cite{Freidel:2014be,Freidel:2015me} 
a unified description of the (dynamical) phase space geometry, which involves the 
symplectic form, the generalized metric (combining the metrics on spacetime and 
momentum space) and the locality metric (which encodes the pairing between 
spacetime and momentum space), satisfying the appropriate compatibility conditions. 
Using such a formulation the relative locality approach can also be introduced in the 
framework of string theory, where it is realized as a generalization to the meta-string theory. 

On the other hand, the nontrivial geometry of phase space has also been considered 
from the perspective of the candidate full quantum gravity theories (and not just 
models of quantum spacetime). For example, curved momentum space can in principle arise 
within the Group Field Theory approach. At the 
elementary level of this framework, spacetime is replaced by a quantum 
field defined on the background of several copies of a certain Lie group 
(associated with relativistic symmetries), which is also a 
curved manifold and whose 
tangent space is given by the related Lie 
algebra. A particle excitation of the field corresponds to 
a single quantum of discrete space(time), while different quanta 
become connected via the combinatorially nonlocal field interactions. Continuous 
spacetime is expected to emerge only in the semiclassical limit and there 
a group field theory should reduce to a certain 
effective field theory (describing perturbations around the classical solution) \cite{Oriti:2012ty}. It is reasonable to suppose that 
the latter field will perceive some effective Lie group as configuration space 
and the (dual) Lie algebra as momentum space. However, one may also envisage 
the opposite situation and then, at the mathematical level, the 
difference between a given group field theory 
and the theory covariant 
under the action of the appropriate Hopf algebra will be the 
combinatorial structure of the former. Indeed, it has been tentatively shown \cite{Girelli:2010fs} that a 
scalar field theory on $\kappa$-Minkowski space is equivalent to the effective, 
semiclassical approximation of a Poincar\'{e} group field theory. Another avenue 
that potentially leads to nontrivial phase space is the already mentioned polymer quantization scheme, 
commonly applied within LQG, which can result in the circular (periodic) momentum space of a quantum mechanical system \cite{Bojowald:2011ge}. 

One might also wonder whether the concept of nontrivial phase space 
should be extended to the domain of field theory, so that it is the phase space of values 
of a given field that has some nonlinear structure (rather than just the phase space on which 
a field is defined). In \cite{Mielczarek:2016ty} (see also \cite{Mielczarek:2016fe,
Trzesniewski:2017fe,Bilski:2017gic}) the authors of this paper proposed the conjecture that ordinary 
field theories are actually the low energy limit of theories whose phase spaces 
of field values are not affine spaces but manifolds with nontrivial geometry or topology. 
This has been called the Nonlinear Field Space Theory (NFST). As a prototypical 
example of such a framework we considered a scalar field theory whose phase space at every point of spacetime, or for every Fourier mode, has 
the symplectic geometry of a sphere (while the background spacetime is assumed to be 
Minkowski space). Performing quantization of the field we then obtain a number of 
predictions typical to the quantum gravity models. Moreover, if the assumed phase space is 
compact, as it is the case for a sphere, the so-called principle of finiteness \cite{Born:1934fy} is automatically imposed on 
a given field, which automatically resolves the ubiquitous problem of UV divergences. 
In \cite{Mielczarek:2017ny} we also preliminarily investigated the application of such a scalar field as the 
matter content of universe in the standard cosmological model, which turned out to lead 
to different interesting results. 

Worth discussing is a further extension of the nonlinear phase space framework to the area of quantum gravity. 
In this context let us take a look at the case of LQG. The starting point of the latter approach is the formalism of 
Ashtekar variables, in which the connection $A$ and densitized triad $E$ are the canonical fields, whose values 
belong to the ${\mathfrak su}(2)$ algebra \cite{Ashtekar:1986yd}. The phase space of classical GR parametrized 
in terms of the Ashtekar variables is still affine. However, passing to LQG, $A$ is subject to the exponentiation
and forms a holonomy, being an element of a compact group ${\rm SU}(2)$ \cite{Rovelli:1989za}. Meanwhile, 
the fluxes constructed with the use of $E$ remain elements of the ${\mathfrak su}(2)$ algebra, which is isomorphic 
to $\mathbb{R}^3$. Therefore, similarly as we observed above in the context of Group Field Theory, 
one can say that the phase space of LQG is partially curved. 

On the other hand, in Sec.~\ref{sec:phases} it was mentioned that the opposite setup for holonomies and fluxes is considered in the recently 
proposed alternative formulation of LQG \cite{Bahr:2015bra}. Namely, $E$ is then subject to the exponentiation, 
while fluxes of $A$ remain elements of ${\mathfrak su}(2)$. The natural next step one could consider is to exponentiate 
both $E$ and $A$, so that the phase space per link of a spin network becomes $\Gamma = {\rm SU}(2) \times {\rm SU}(2)$, 
which is a compact manifold. The idea of generalizing LQG to such a case was discussed in \cite{Rovelli:2015fwa}. 
While, so far, it has been mostly analyzed only in the (2+1)-dimensional theory, significant progress 
towards the extension of this framework to 3+1 dimensions has recently been achieved \cite{Dittrich:2017nmq,Dittrich:2016typ,Riello:2017iti}.

\section{Synthesis: the first map} \label{finalmap}

\begin{figure}
\begin{center}
\includegraphics[width=9cm]{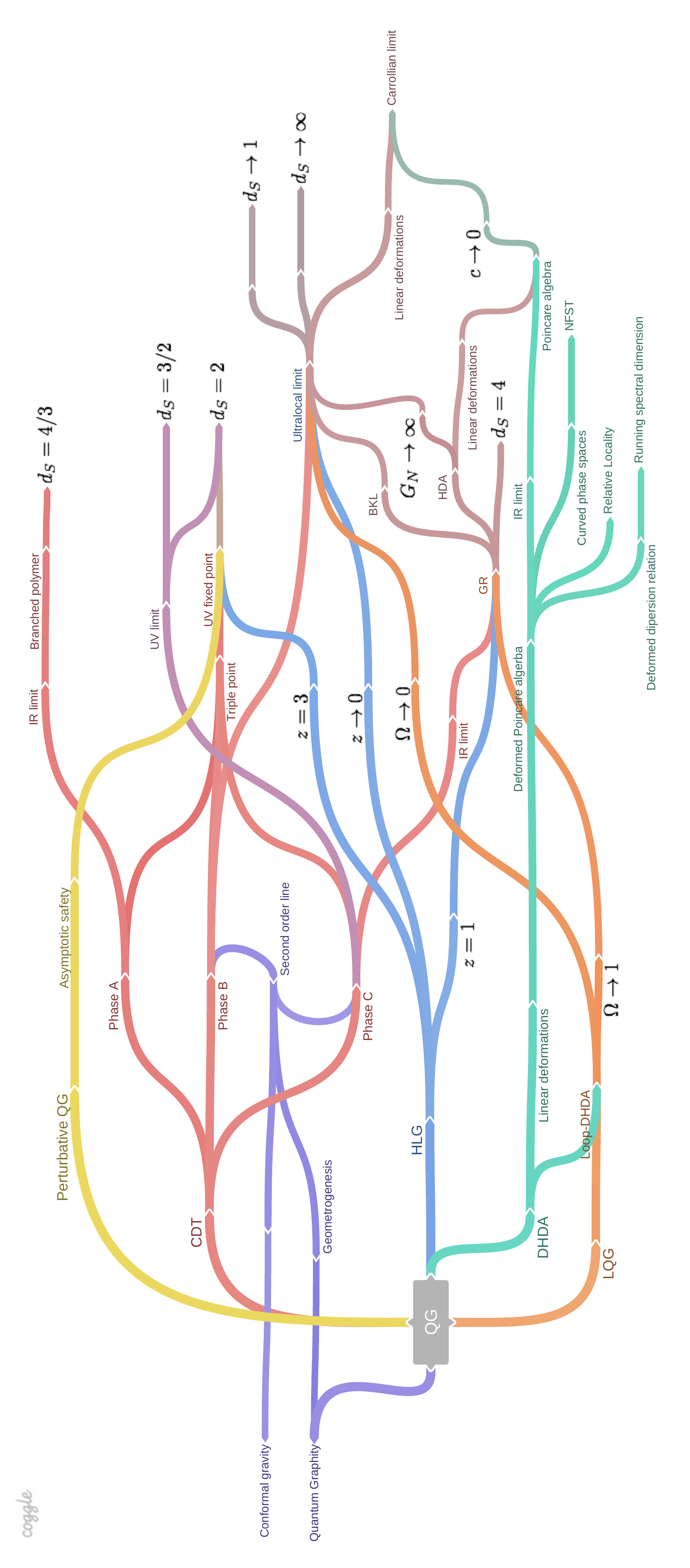} 
\end{center}
\caption{Map of relations between some aspects of quantum gravity, constructed on the basis 
of the architecture presented in Fig.~\ref{Fig2}.}
\label{FinalMap}
\end{figure}

In this section we make the first attempt to collect the results discussed above 
into a single map, representing relations between different approaches to quantum gravity. 
The map in Fig.~\ref{FinalMap} has been constructed using the architectural structure exemplified in Fig.~\ref{Fig2}. 
Its core elements are:

\begin{itemize}
\item Quantum deformations of HDA and 
the corresponding deformations of the Poincar\'{e} algebra. The bridge between them allows to relate models 
of quantum spacetime with the candidate theories of quantum gravity. 
\item Various phases of the gravitational field, invariant under different symmetries, 
which are described by the (deformed) symmetry algebras. 
\item The phase transition lines, meeting at the triple/quadruple point. This point is possibly related 
to a UV fixed point of the RG flow, which is also supposed to coincide with the Lifshitz point of HLG. 
The asymptotic safety conjecture is expected to be realized there.  
\item Nonlinear phase spaces of particles and fields, which can lead to deformations of the relativistic symmetries. 
\item The ultralocal state, predicted to occur in such approaches to quantum gravity as the effective LQG, CDT and 
Ho\v{r}ava-Lifshitz gravity.
\end{itemize}

\section{Summary and discussion}

In this paper, in the spirit of the architectural approach to a physical theory, 
we carried out our first attempt to draw a mind map of quantum 
gravity. It is based on the most up to date results, obtained within various 
approaches to the Planck scale physics. 

As several researchers have observed (see Sec.~\ref{sec:phases}), there are strong reasons to suppose that quantum gravity belongs to the realm of complex systems. 
A crucial indication is that the gravitational field is a system with a huge (perhaps infinite) number of degrees 
of freedom, which are subject to strongly non-linear interactions. Systems of this 
kind exhibit various emergent phenomena, such as the nontrivial phase structure, containing 
phase transitions. Consequently, presumably there is no single effective description of states of 
quantum gravity but rather several ``equations of state'', corresponding to 
different phases. Nonlinearity implies the existence of various underlying mechanisms, 
each deserving a separate analysis and explanation with the help of adequate 
variables and a set of concepts. Similarly, the functioning of a human brain 
cannot be described using a single equation. Rather than that, a huge network of biochemical 
processes, composed of numerous contributing reactions, has to be considered. 

The concept of a network, or graph, occurs naturally in the description 
of complex systems. It consists of two essential levels: \emph{structural} and 
\emph{functional}. The first one means that a graph serves as a method for 
capturing relations between elementary constituents of the complex system 
(e.g. connections between neurons in a brain) in the visual language. 
Meanwhile, at the second level a graph is a way of representing different 
operating processes and emergent phenomena in the system 
(e.g. epileptic seizures happening in a brain). 

While the first (structural) point of view is extensively used in the investigations of 
quantum theory of gravity (e.g. spin networks, Regge calculus, tensor networks), the 
second one (functional) has had only a residual appearance in the quantum 
gravity research so far. This paper constitutes the first serious attempt to introduce the 
functional network approach into the domain of quantum gravity. In our discussion we have considered 
a simple architecture, combining such issues as the gravitational phase structure, running dimension 
of spacetime, asymptotic safety of the theory as well as deformations of relativistic symmetries and 
nonlinear phase space structures. The obtained construction indicates certain 
missing points, which lead to a number of open questions listed in the Appendix.    

Among the most important observations that contributed to the structure of the presented map 
we should mention:  
\begin{itemize}
\item Different phases of quantum gravity will have different effective descriptions 
(equations of state). 
\item There may be a direct connection between the UV fixed point in the 
asymptotic safety scenario, the Lifshitz point in the Ho\v{r}ava-Lifshitz gravity 
and triple (quadruple) point on the CDT phase diagram. 
\item A variety of approaches to quantum gravity predict the dimensional reduction 
of spacetime at small scales. 
\item A given deformed hypersurface deformation algebra in the linear limit 
leads to the corresponding deformation of the Poincar\'{e} algebra. 
\item The nonlinear structure of phase space is associated with deformations 
of general and special relativistic symmetries. 
\item The ultralocal (silent) state probably plays an important role in quantum gravity. 
\end{itemize}

Quantum gravity is a fascinating jigsaw puzzle. Many of its elements are definitely 
still missing. The only way to find out which elements we need to look for is to put 
together the pieces that are already available. This has been the most important motivation behind 
our proposal, which deserves a further systematic extension, including additional 
models and aspects of quantum gravity.

\ack

This work is supported by the Iuventus Plus grant No.~0302/IP3/2015/73 from the Polish Ministry 
of Science and Higher Education. TT was additionally supported by the National Science Centre 
Poland, project 2014/13/B/ST2/04043.

\section*{Appendix -- Open problems}

Here we collect the main open problems that emerged in the course of our investigations:

\begin{enumerate}
\item What is the behavior of the spectral dimension at the triple point on the phase diagram of CDT?
\item In the continuum, is the CDT triple point a Lifshitz point with $z = 3$?
\item Is this point also a nontrivial UV fixed point?
\item What is the order of the $A-B$ phase transition in CDT?
\item Is the BKL conjecture realized in the phase $B$ of CDT?
\item What is the nature of the Ho\v{r}ava-Lifshitz theory with the critical exponent $z \rightarrow \infty$?
\item What are the properties of the Ho\v{r}ava-Lifshitz theory with $z = 6$, which leads to $d_S = \frac{3}{2}$?
\item What are the properties of the Ho\v{r}ava-Lifshitz theory with $z = 9$, which leads to $d_S = \frac{4}{3}$?
\item What is the (class of) DHDA corresponding to the $\kappa$-Poincar\'{e} algebra?
\item Is the loop-deformed Poincar\'{e} algebra associated with certain nonlinear structure of phase space?
\end{enumerate}

Finding answers to the above questions will be a significant guidance in the further extension of the 
map of quantum gravity.

\end{document}